\documentclass{aa}
\usepackage{natbib}
\usepackage{epsfig}
\usepackage{csquotes}
\usepackage{tikz}
\usetikzlibrary{decorations.pathmorphing}
\usetikzlibrary{decorations.pathmorphing,calc}
\def\ut#1{\mathop{\vtop{\ialign{##\crcr
     $\hfil\displaystyle{#1}\hfil$\crcr\noalign
     {\kern1pt\nointerlineskip}\hbox{$\hfil\sim\hfil$}\crcr
     \noalign{\kern1pt}}}}}

\def\undersymbol#1#2{\mathop{\vtop{\ialign{##\crcr
     $\hfil\displaystyle{#2}\hfil$\crcr\noalign
     {\kern1pt\nointerlineskip}\hbox{$\hfil#1\hfil$}\crcr
     \noalign{\kern1pt}}}}}

\usepackage{graphicx}

\begin{document}

\title{Cosmic voids and the kinetic analysis. \\ III. Hubble tension and structure formation in the late Universe}
       \author{V.G.Gurzadyan\inst{1,2}, N.N.Fimin\inst{3}, V.M.Chechetkin\inst{3,4} }

              \institute{Center for Cosmology and Astrophysics, Alikhanian National
Laboratory, Alikhanian Brothers str.2, 0036, and Yerevan State University, Manukian str.1, 0025 Yerevan, Armenia \and
SIA, Sapienza Universita di Roma, Via Salaria 851/881, 00191 Rome, Italy \and Keldysh Institute of Applied Mathematics of RAS, Miusskaya Sq. 4, 125047 Moscow, Russia \and Institute of Computer Aided Design of RAS, 2nd Brestskaya st., 123056 Moscow, Russia}

   \offprints{V.G. Gurzadyan, \email{gurzadyan@yerphi.am}}
   \date{Submitted: XXX; Accepted: XXX}

 \abstract{We study structure formation in the late Universe within the Vlasov kinetic self-consistent field approach. Our work is principally focused on the use of the modified gravitational potential with a repulsive term of the cosmological constant, which is directly linked to observations that enable characterizations of the Hubble tension as the result of local  and global flows. We formulate the criteria for the formation of the semi-periodic gravitating structures, along with the predictions of their quantitative scales associated with observable parameters. Our principal conclusion is that filament formation in the Local (late) Universe can proceed as a deterministic process that is distinct from the structures at larger scales that result from the essentially stochastic dynamics of density perturbations.     
}


   \authorrunning{V.G. Gurzadyan, N.N.Fimin, V.M.Chechetkin}
   \titlerunning{Cosmic voids and kinetic analysis. III.}
   \maketitle
%

\section{Introduction}

The observational indications for the accelerated expansion of the Universe and the positive cosmological constant have essential influence on the theoretical studies of the various phases of the cosmological evolution. The Hubble tension \citep{Verd,R,R1,Val,Dai1,Dai2}, as the plausible tension between the late and early Universe, has also stipulated the consideration of theories and models  (e.g.\citep{photon,Bo,Baj}) addressing a broad spectrum of relevant issues, including the formation of cosmic structures. 

Zeldovich \citep{Z,SZ,SS} pancake theory predicts the formation of cosmic structures based on the evolution of initial density perturbations \citep{Peeb}, with the essential use of the Lagrangian singularity theory. The pancake theory originated from cosmology thus linked the "hydrodynamics" of the Universe to symplectic geometry \citep{Arn,ArnP}. The observational facts on the non-zero cosmological constant or the possible tension of the late (local) and early (global) Universe would obviously also have an influence on structure formation and pancake theory.   

The concept of two Hubble flows, the local and global one, with a non-identical Hubble constant, is among the suggested approaches to deal with the Hubble tension and the dark sector \citep{GS7,GS8}. That approach is based on the theorem proved in \citep{G1}, positing that the general function for the force satisfying the sphere-point mass gravity's identity takes the following form: 
\begin{equation}
F=-{\frac{GMm}{r^{2}}}+{\frac{\Lambda c^{2}mr}{3}}.
\end{equation}
This function does not satisfy the second statement of Newton's shell theorem, namely, in this case the gravitational field inside a shell is no more force-free. This formula, that is, taking as the second term the cosmological constant in weak-field general relativity and McCrea-Milne non-relativistic cosmology \citep{MM} (see also \citep{Z81}) enables
the description of the dynamics of groups and clusters of galaxies, as well as other phenomena \citep{GS2,G2,GS3,GS4}. We also note that there are observational indications \citep{Kr} for the influence of the halo on the spiral galaxy structure, thus supporting the non-force-free field inside a shell predicted by Eq. 1.   

Our present study is a continuation of previous works \citep{GFC1,GFC2}, based on the use of the Vlasov kinetic technique applied to gravitating systems and revealing the paths of the formation of the filamentary cosmic structures. The question of the ordered relaxation of perturbations is rather subtle in the system along the selected direction, such as for filaments, along two mutually perpendicular directions. It is clear that the local structure formation essentially differs from the hydrodynamics at large scales, especially regarding the averaging of a self-consistent gravitational field in a system of particles. We show that the deterministic way for emergence of large-scale structures,
coherent in the sense of having a translational invariance of substructures, can be crucial in the Local Universe. This approach differs from  ``order out of chaos'' scenario of the pancake theory at global cosmological scales but is complementary to it at local scales. The role of the repulsive second term of Eq. 1 is crucial and also determines  the scales of  structure formation and sizes of the voids, thus linking the predictions to observations \citep{BH} (and references therein).

\section{Integral form of the Poisson-Liouville-Gelfand equation for the system of
 gravitating particles}
 
The system of Vlasov-Poisson equations for describing the cosmological dynamics
in a system of $N$ particles of identical masses $m_i =m\equiv 1$
 can be represented in the following form. We note that we implicitly
assume that the system is considered in the domain
$\Omega \subseteq {\mathbb R}^2$
space with a smooth boundary $\partial \Omega$,
where ${\rm {diam } \:\Omega}\le \infty$, namely, the size of the region can be continued up
to infinity. This system is expressed as follows:\ 
$$
\frac {\partial F({\bf x},{\bf v},t)}{\partial t} +
{\rm {div}}_{\bf x}({\bf v}F)+\widehat {G}(F;F)=0,
$$
$$
\widehat{G }(F; F) \equiv -\big ({\rm{div}}_{\bf v}F\big )
\big({\nabla }_{\bf x}(\Phi (F)\big),
\eqno{(2)}
$$
$$
{\Delta}_{\bf x}^{(3)}\Phi(F)\big|_{t=t_*,\forall {t_*}\in\mathbb{T}}=
AS_3 G \int F( {\bf x},{\bf v},t_*)\:d{\bf v} - \frac{c^2\Lambda}{2},
\eqno{(3)}
$$
$$
S_3\equiv{\rm{meas} \:{\mathcal S}^2} =4\pi,~~{\mathcal S}^d=\{{\bf x}\in
\mathbb {R}^d, |{ \bf x}=1|\},
$$
where $ F({\bf x},{\bf v},t)$ is the distribution function of gravitationally interacting
particles,
$A$ is a normalization factor for particle density, and $t_{*}$ is a fixed
instant of time.
Equation 3 is the Poisson equation for the potential of Eq. 1, taking into account the repulsive cosmological term.

The third term in the r.h.s. of the kinetic equation (Eq. 2) can be represented as
a ``source-like '' form
$$
\widehat{G}(F; F) =  {\bf G}(F)\frac{\partial F}{\partial {\bf v}},~~~
{\bf G}(F)=-\nabla_{\bf x}\Phi(F),
\eqno{(4)}
$$
$$
\Phi(F)= AS_3 G \int\int  {\mathfrak K}_3
({\bf x}-{\bf x}')  F({\bf x}',{\bf v}',t_*)\:d{\bf x}'d{\bf v}' +
$$
$$
\frac{\Lambda c^2}{12}|{\bf x}|^2 + \widehat{\mathfrak B}_3 ({\bf x},{\bf x}'),~~
{\mathfrak K}_3 ({\bf x}-{\bf x}')=-{|{\bf x}-{\bf x}'|^{-1}},
$$
\noindent
where 
$\widehat{\mathfrak B}_3 ({\bf x},{\bf x}')$
is an operator term that takes into account
border influence
domain (for $d=2,$ the situation is similar, but
${\mathfrak K}_2 ({\bf x},{\bf x}')=\ln\:{|{\bf x}-{\bf x}'|}$, $S_2=2$).
The Newtonian potential
$\Phi_{N}(r)= -G m/r$ increases
on an interval of $r\in (0,+ \infty )$ ($\Phi_N \in (-\infty,0)$),
while the potential of Eq. 1:
$$
\Phi_{GN}(r)\equiv -G M/ r - \frac{1}{2}c^2\Lambda r^2,
$$
has a maximum of 
$$
\Phi_{GN}^{(max )}=
-\frac{1}{2}G(3Mc^{2/3})\Lambda^{1/3},
$$
at the critical radius
$$
r_{crit}= \big(3G M/\Lambda c^2\big)^{1/3},
\eqno {(5)}
$$
namely, the case where the $\Lambda$-term  starts to dominate in the l.h.s. of Eq. 1.

The radius, $r_{crit}$, determines the mutual contribution of the gravitational attraction and repulsive $\Lambda$-term. That is to say, at relevant scales, it defines the local Hubble flow given by the equation:
$$
H_{local}^2 = \frac{8 \pi G \rho_{local}}{3} + \frac{\Lambda c^2}{3},
$$
where $H_{local}$ and $\rho_{local}$ are the local values of the Hubble constant and the matter density, respectively \citep{GS8}. The critical radius, $r_{crit}$, also defines the scaling between the semi-periodic coherent structures (i.e., the voids) formed as a result of action of attraction and repulsion \citep{GFC1,GFC2}.

We go on to consider the stationary dynamics, thus we set
$F= F( {\bf x},{\bf v})$. The subsequent analysis mainly concerns
the second equation of the system Eqs. 2-3, which is the Poisson equation
relative to the potential with no explicit time dependence.
That is why varying the Hilbert-Einstein-Poisson action
 \citep{VFC2,Ved} 
includes a separate variation of fields (for a fixed particle distribution) and the distribution function. 
Thus, the considered below approach is applicable
for adiabatic processes at quasi-equilibrium (weakly varying)
particle distribution functions.
In this case, for the distribution function, we can use the energy
substitution \citep{Ved}, namely: $F({\bf x},{\bf v})=f(\varepsilon)
\in C^1 _{+} ({\mathbb R}^1)$, where
$\varepsilon = {\bf v}^2/2 + \Phi ({\bf x})$.

Thus, the particle density on the right side of the Poisson equation
can be expressed in terms of the integral of the known equilibrium solution
of Vlasov equations,
having the most transparent physical meaning, namely, that of
Maxwell--Boltzmann distributions $f=f_0(\varepsilon )=
AN\exp(-\varepsilon /\theta),$
$$
\Delta \Phi({\bf x}) = AN G_3 S_3^2
\bigg (\int_{y\in [0,\infty ]}\exp \big(-y^2/(2\theta) \big)y^2\:dy
\bigg )\cdot 
$$
$$
\exp(-\Phi/\theta)-\frac{c^2\Lambda}{2},~~~
A,\theta,R_\Omega \in {\mathbb R}^1,
\eqno {(6)}
$$\noindent
where  $R_\Omega$ is the radius of the region $\Omega$ as 
a ball in accordance with the 
Gidas-Ni-Nirenberg theorem \citep{Gi,Du}. Here,
$\theta $ has the meaning of dynamic kinetic temperature
of the system of interacting
particles. The meaning of the last value is by no means obvious,
as thermodynamic equilibrium is globally absent here.

Introducing the (kinetic) temperature according to the general definition \citep{V1,V2}, 
 $\theta\equiv -
({\partial \Phi/\partial x_j })/({\partial \rho/\partial x_j })
\big|_{j=\overline{1,d}}$, we get its
 dependence on the local properties of the self--consistent potential. We note that the temperature, generally speaking,
can be considered as a tensor quantity, but for simplicity,
we confine ourselves to the scalar temperature in the above
formula, implying summation over the repeated indices.

Thus, the Poisson equation (Eq. 3) takes the form of an inhomogeneous 
Liouville--Gelfand (LG) equation
\citep{Gelfand,Du} with local (generalized) temperature changing the sign 
depending on the value of the derivative of the potential at a given point.
As mentioned above, for
two-particle problem --- in particular, for a formal pair of a central
coalescence of the main part of the particles and the conditional ``far'' particle
(generalized Milne--McCrea model), a dominant one can become the
repulsive force due to
the presence of a quadratic term in the radius-vector containing
cosmological parameter of Eq 1. 
While the generalized
indefinite thermodynamics of a system of gravitating particles becomes
similar to that for the Onsager vortices in the classical
hydrodynamics \citep{FM}
in the context of the formation of quasi-ordered large--scale coherent
structures). The existence of solutions of the LG equation for large system sizes
supports (unlike the case $\theta >0$ for a non--positive
Newton's gravitational attractive potential)
the existence of solutions to the Vlasov equation. In this case, the 
gaps in the solution due to the fact that for $\Phi_{r}' \gtrless 0$
solutions can exist on non-intersecting intervals (here
variable $r= |{\bf x}|$)
on both sides of the maximum of the self-consistent potential,
corresponding to $\Phi_{GN}$ for the two-particle potential).
This can be shown using the parametric Young's inequality
\citep{Pokh}. To do this, we multiply both sides
of non-homogeneous LG equation to the expression $q_1(\Phi )q_2({\bf x})$, 
where $q_1\in C^1({\mathbb R}^1)$ is such that 
$Q_1\equiv -\big( 1+\exp (\Phi /\theta ) \big )^{-1}=\int_0^\Phi q_1(z) dz$,
and the test function 
$q_2 ({\bf x })\in C^2({\mathbb R}^d)$ can be represented 
as: 
$$
q_2 ({\bf x})= q_2 ({\bf y}), ~~{\bf y}={\bf x}/R (0\le R \le R_\Omega/2),
$$
$$
q_2 ({\bf y})=1~ \mbox{\rm {for}}~ |{{\bf y}}|\le 1, ~~q_2 ({\bf y})=0 
~ \mbox{\rm {for}}~ |{{\bf y} }|\ge 2, 
$$
$$
q_0 
\equiv \int \big ( |\Delta q_2|^2/q_2 \big )d{\bf y}<\infty. 
$$ 
Then, we integrate this and using the Young inequality obtain the expression:
$$
\int_\Omega |\nabla \Phi |^2 \frac {\exp (\Phi /\theta ) q_2({\bf x}) 
\big ( 1 - \exp (\Phi /\theta ) \big ) }{\big (1 + \exp (\Phi /\theta )\big )^3 \theta^2}\:
d{\bf x}
$$
$$
\:\le \:
-\int_\Omega \frac{\lambda}{2\theta \big( 1+\exp(\Phi/\theta) \big)^2}d{\bf x}
+
\eqno {(7)}
$$
$$
+\frac{q_0 \theta}{2 R \lambda} - \int_\Omega \frac{c^2\Lambda}{2} 
q_1(\Phi)q_2({\bf x}) d{\bf x}
\:\le\:-\frac{\lambda S_3}{2\theta}R^3 + \frac{q_0 \theta}{2\lambda R}+ 
\frac{2c^2\Lambda}{3 \theta}R^3,
$$
$$
q_0 \equiv \int \big ( |\Delta q_2|^2/q_2 \big )d{\bf y}<\infty,
$$\noindent
where 
$$
\lambda \equiv AN \gamma_3 S_3^2 J(\theta),~~~ J(\theta)
\equiv \int_0^{v_{max}} \exp\big(-v^2/(2\theta)\big)v^2dv.
$$ 
\noindent
The positivity condition
on the left-hand-side gives us the parameter link: for $\theta \gtrless 0$
the compatibility condition is $c^2\Lambda \gtrless 3\pi \lambda $.

Therefore, if there are appropriate
relations between the parameters of the problem, the system of Vlasov--Poisson equations
has solutions of the type of distribution functions that admit the energy
substitution. The potentials of the
gravitational field, which have the property of convexity (cf. the difference with tidal fields \cite{GO}) in the general case,
for an arbitrary $R_\Omega \le \infty $ (in contrast to the case of the attraction potential),
 there is a limitation
$R_\Omega < \big (q_0 \theta^2 / (4\pi \lambda^2) \big )^{1/4}$.

Now we go on to examine in detail the properties of the gravitational potential and
the influence of the cosmological term, as well as its influence
at a given point of boundary conditions for
region of space $\Omega$.
As already noted in \citep{GFC1,GFC2},
in the formulation of the Dirichlet problem for the Poisson equation with a constant 
right-hand side
on the boundary of the $\Omega$ region, according to
averaging gravitational field
outside the compact subdomain $\Omega_0$ containing
system of particles and located inside
domain $\Omega$ (${\rm meas }\,{\Omega_0}\ll {\rm meas }\,{\Omega}$), we can
assume that the data on the $\partial \Omega$ boundary is given by
in accordance with the theorem in \cite{G1}. Then we consider what
happens in the context of other similar equations.

We should first consider the properties of the boundary value problem for
Eq. 6 in the following:
$$
\Delta\phi +\lambda \exp (\phi )= \frac {c^2\Lambda }{2\theta },~~
\phi \equiv -\frac{\Phi}{\theta},~~~\phi ({\bf x})\big |_{{\bf x} \in \partial \Omega }=
\phi_b (R_\Omega).
\eqno {(8)}
$$
If we assume a priori the quantity $\phi $ to be small in norm, we obtain from this a linear
equation; in principle, a linearization near any norm-finite solution
$\phi_0$
homogeneous equation gives a similar result, although with the change
$\lambda \to \lambda \exp (\phi_ 0);$  for simplicity, we further assume
$\phi_0=0$): $\Delta\phi+\lambda \phi={c^2\Lambda}/({2\theta}).    $
We note that the boundary condition at $r\to R_\Omega$ must take the form:
$\sim k_1 (\theta)/R_\Omega+k_2(\theta)$).

Thus, we have the Dirichlet problem for the inhomogeneous Helmholtz equation, the solution
of which in a three-dimensional radially-symmetric
case takes the form 
$$
\phi_3({\bf x })\sim r^{-1} \big (\cos (\sqrt {\lambda }r)+
\sin (\sqrt {\lambda }r )\big)\big |_{r =|{\bf x}|}+{c^2\Lambda }/({2\theta }\lambda ),
\eqno {(9)}
$$
\noindent
which is analogous to the consideration of the inhomogeneous Poisson equation \citep{G1} and an additional
condition, $\phi ( 0)<\infty,$ can be satisfied by a discrete or locally finite
density of mass carriers $\rho $, namely,
$\exists \omega : \omega = \{ 0\le r_{-}\le R_{ min }\ll R_\Omega\}:\:\rho (r_{-})=0$).
We note that the Dirichlet condition is consistent with the fundamental system of solutions
of the considered equation. For the stability of the solution of the boundary value problem, and
to set the corresponding condition for the case $\lambda \equiv 0$
(for a non-homogeneous Poisson equation), the Dirichlet condition on $\partial \Omega$
is expressed as:
$$
\phi_3 \big( |{ \bf x}|\to R_\Omega  \big)\sim R_\Omega^{-1}
+{c^2\Lambda }R _\Omega^{2}/(2\theta ).
\eqno {(10)}
$$ 
Obviously,
we can pose the asymptotic conditions at infinity for a linearized model reducing
it to the Helmholtz equation, in accordance with the known
Sommerfeld radiation conditions \citep{Vladimirov}. At the same time, for the 
Poisson equation with the right-hand side, this is impossible, because
$\lim_{r \to\infty }| \phi(r)|\to\infty$. However, in a formal sense,
for the last equation, it is possible to associate solutions
of exact hydrodynamic consequence of the system of Vlasov--Poisson equations with
Milne-McCrea model. Hence, the latter model
should be seen as a solution
the Dirichlet problem in a ball with a large but finite radius $R_\Omega$; moreover,
the possibility of setting periodic boundary conditions from the point of view of physical
problem statement is
highly doubtful. The Helmholtz equation can be solved on a semi-infinite interval
(moreover, periodic conditions may well be set), and it corresponds to
non-relativistic cosmological model with pseudo-periodic decrease of density amplitude.
We note, in fact, that these linear models are hierarchical steps for a more general
models with a conditionally equilibrium distribution of particles in space.

For a more detailed analysis of the situation,
it is expedient to pass to the integral representation of the equation for
gravitational potential. The equation for the potential with
Maxwell--Boltzmann particle distribution, corresponding to
internal Dirichlet problem in a bounded domain $\Omega$
(under boundary conditions corresponding to the McCrea-Milne model), takes the following form:
$$
\Phi ({\bf x}) = \lambda_I \int_{\Omega'}{\mathcal K}({\bf x},{\bf x}')
 \exp \big( -\Phi ({\bf x}')/\theta \big) d{\bf x}' -
\frac{c^2\Lambda}{12}{\bf x}^2 + C_0,
\eqno{(11)}
$$
$$
{\mathcal G}({\bf x},{\bf x}') \equiv 4\pi \sum^\infty_{\ell =0}
\sum_{m=-\ell}^\ell \frac{Y_{\ell m}^{*} (\vartheta',\varphi')
Y_{\ell m} (\vartheta,\varphi)}{2\ell + 1}
\frac{x_{<}^\ell x_{>}^\ell}{{{R}}^{2\ell + 1}_\Omega},
$$
$$
x_{<}={\rm{min}}(|{\bf x}|,|{\bf x}'|),
$$
$$
x_{>}={\rm{max}}(| {\bf x}|,|{\bf x}'|),
$$
$$
{\mathcal K}(|{\bf x}-{\bf x}'|)\equiv {\mathcal G}({\bf x},{\bf x}') -
\frac{1}{|{\bf x}-{\bf x}'|},~~~
C_0 = -\frac{\gamma_3 Nm}{{{R}_\Omega}} - \frac{c^2\Lambda {{R}_\Omega}^2}{12},
$$
$$
\lambda_I =\lambda /S_3.
$$

In essence, the above is the explicit form of the equation for the potential introduced in the
expression (Eq. 3), where ${\mathcal G}( {\bf x},{\bf x}')$ is the Green function
for the inner boundary value problem in the domain $\Omega$; in this case, due to
symmetry of the latter, we have:
$$\int_{\Omega'} {\mathcal G}( {\bf x},{\bf x}')\rho ({\bf x}')d{\bf x}'\to
C_1 (={\rm {const}) }.
\eqno{(12)}
$$

 We go on to introduce a new variable $ U( {\bf x})\equiv (\Phi ({\bf x})-C_0)/\theta,$
for a symmetric problem, where $C_1$ will enter the constant $C_0$,
the above equation
can be written as the Hammerstein integral equation of the form:
$$
U({\bf x})=
\widetilde{\lambda}  \widehat{\mathfrak G}({U})+
\alpha(\theta,\Lambda){\bf x}^2,
$$
$$
\widehat{\mathfrak G}({U})\equiv \int_{\Omega'} {\mathcal K}(|{\bf x}-{\bf x}'|)
\exp\big(-{U}({\bf x}')\big)d{{\bf x}'},
\eqno{(13)}
$$
$$
\widetilde{\lambda} \equiv \widetilde{\lambda_{II}}(\theta) 
\equiv \frac{\lambda_{I}}{\theta}\exp(-C_0/\theta),~~~
\alpha(\theta,\Lambda)= -\frac{\Lambda c^2}{12\theta}.
$$

To clarify the basic properties of the solution of the potential equation, first, we
assume that the  value of the potential differs 
little from the constant value ${U}_0$, namely,\:
${U}({\bf x})={U}_0-\phi ({\bf x})$, $|{\phi }|\ll {U}_0$;
from a physical point of view,
we consider the region near the inflection point of the potential,
where for multi-particle systems there should be an equilibrium plateau
forces of attraction and repulsion (unless, of course,
the size of the $\Omega$ system is large enough). A linearization of the equation near
(quasi)constant solution ${U}_0$ leads to the homogeneous
Fredholm second kind equation
$$
\phi ( {\bf x})+{\lambda}_{II}\int_{\Omega'}
{{\mathcal K }}( |{\bf x}-{\bf x}'|)\phi ({\bf x}')d{\bf x}'=0,
$$
$$
\lambda_{II} = \widetilde{\lambda_{II}} (\theta)\exp(-{U}_0).
\eqno{(14)}
$$
In addition to the trivial solution, this equation has
a set of periodic (in the 3D space) solutions.
We introduce eigenfunctions of symmetrical kernel ${{\mathcal K }}( |{\bf x}-{\bf x}'|)$,
and consider rep\-re\-sen\-ta\-ti\-on of solution in the form of Fourier series
$$
\phi ({\bf x })= \sum_{\bf n}\zeta_{\bf n}
\exp \big ( i {\bf n}{\bf x}),~~~
{\bf n}= {\rm{colon}}( n_i )_{i=1,2,3},
\eqno{(15)}
$$
\noindent
where $n_i =2\pi /T_3$, $T_3$ is a spatial period.
For the 1D case and 
$\Omega=\{r<R_{\Omega}\},$ we have: 
$$
\phi_n (r)=\sqrt{2/R_\Omega}
\sin\big( \pi n r/R_\Omega \big),
$$
$$
\lambda_{n}^{-1}=4\pi n^{-2}\big(1-\cos(nR_\Omega)\big),~~
\lim_{R_\Omega\to \infty}\lambda_{n}^{-1}=
4\pi/n^2.
\eqno{(16)}
$$

We substitute this expression into the linearized equation and obtain:
$$
{\lambda}_ {II}\int_{\Omega'}
{\mathcal K}( |{\bf x}-{\bf x}'|)\exp \big (-i {\bf n}({\bf x}-{\bf x}')\big )
d( {\bf x}-{\bf x}')=-1.
\eqno{(17)}
$$
The criterion for the existence of periodic solutions is:
$$
\lambda_{II}\le \lambda_{II}^{crit},~~~
1/\lambda_{II}= 
-S_3\int^{R_{\Omega}}_0 {\mathcal K}(y)\cdot \big(\sin ( ny )/n\big) ydy,
$$
$$
1/\lambda_{II}^{crit} \equiv -S_3\int^{R_{\Omega}}_0 {\mathcal K}(y)y^2dy,
\eqno{(18)},
$$
\noindent
the latter critical value corresponds to the $T_3\to \infty$ case.
For $\lambda_{II}<\lambda_{II}^{crit}$, the solutions do not become periodic.
We can represent the condition of spatial periodicity of solutions in terms of
``critical temperature," namely:
$$
S_3 \lambda_I \theta^{-1}_{crit} \exp (-C_0\theta^{-1}_{crit})\int^{R_{\Omega}}_0
 |{ \mathcal K}(y)|y^2dy=1,
 \eqno{(19)}
$$\noindent
whose nontrivial solution is unique and is expressed in terms of
transcendental  Lambert $W$--function.
It should be noted that with a formally negative kinetic
temperature criterion is satisfied for the prevailing repulsive forces in the system,
that is, over very long distances. Thus, in principle,  
two types of periodicity of large structures exist -- for the potential proper
attraction at positive temperatures and for the repulsion potential
for negative temperatures.

The linear equation (Eq. 14) above does not contain any inhomogeneity.
However, it does carry information about the Dirichlet conditions and contains
Green's function associated with the original Poisson equation. As the next step, we
can consider
inhomogeneous linear equation taking into account quadratic repulsion
 $\alpha (\theta,\Lambda ){\bf x}^2$.

For this purpose, it is necessary to consider a
linearization of Eq. 13 with selection of the inhomogeneous term: 
$U({\bf x})=U_0-\alpha |{\bf x}|^2 +\phi({\bf x})$.
  In this case, we obtain a linear equation:
  $$
  \phi({\bf x})=\lambda^\dag \int_{\Omega'}{\mathcal K}(|{\bf x}-{\bf x}'|) 
  \phi({\bf x}')d{\bf x}' +\widehat{\beta}({\bf x}),
   \eqno{(20)}
  $$
  $$
  \widehat{\beta}({\bf x})\equiv  -\lambda^\dag\int_{\Omega'}  
  {\mathcal K}(|{\bf x}-{\bf x}'|)\alpha |{\bf x}'|^2 d{\bf x}' -\alpha |{\bf x}|^2,~~~
  \lambda^\dag\equiv-\widetilde{\lambda}\exp(-U_0).
  $$
Obviously, we then get an inhomogeneous
Fredholm--type equation with weak polar kernel ${\mathcal K}(|{\bf x}-{\bf x}'|)$;
this is in accordance with \citep{Vas},
who introduced a new function $g({\bf x})\equiv \phi({\bf x})-\widehat\beta ({\bf x})$. By construction it
is sourcewise representable in terms of the kernel ${\mathcal K}$. Therefore, according to
the Hilbert--Schmidt theorem \citep{Tricomi},
the function $g({\bf x})$ can be expanded into a Fourier series in terms of the eigenfunctions of the
 mentioned kernel:
$$
g({\bf x})= \sum_{\bf n}g_{\bf n} \exp\big(  -i{\bf n} {\bf x} \big),
$$
$$
g_{\bf n} = \langle \widehat\beta ({\bf x}),  \phi_{\bf n}({\bf x}) \rangle 
\equiv\int_\Omega \widehat\beta ({\bf x}) \exp\big(  -i{\bf n} {\bf x} \big).
 \eqno{(21)}
$$
\noindent
Since
$$
g_{\bf n}= \frac{1}{\lambda_{\bf n}} \langle \phi ({\bf x}),  
$$
$$
\phi_{\bf n}({\bf x}) \rangle =
\frac{\lambda^{\dag}}{\lambda_{\bf n}} \langle \phi ({\bf x})-\widehat\beta ({\bf x}),  
\phi_{\bf n}({\bf x}) \rangle, 
 \eqno{(22)}
$$
\noindent
then if the solution $\phi({\bf x})$ exists, then (for $\lambda^{\dag}\neq\lambda_{\bf n}$)
$$
\phi({\bf x}) = \sum_{\bf n}g_{\bf n}\phi_{\bf n}({\bf x})-\widehat\beta ({\bf x})=
-\widehat\beta ({\bf x})+
\lambda^{\dag}\sum_{\bf n}\frac{\langle
-\widehat\beta ({\bf x}),\phi_{\bf n}
 \rangle \phi_{\bf n}({\bf x})}{\lambda_{\bf n}-\lambda^{\dag}}.
 \eqno{(23)}
$$ 

The solution in the form of a series (23) was obtained under the assumption that it exists. 
Let's check that
expression (Eq. 23) is indeed a solution, that is, it satisfies Eq. 20, with the
right side $-\widehat\beta ({\bf x})$; to do this, we substitute into this inhomogeneous equation (Eq. 23):
$$
\lambda^{\dag} \widehat{\mathfrak G}'|_{U=U_0}\phi =\lambda^{\dag} \widehat{\mathfrak G}'|_{U=U_0}
\big(-\widehat\beta ({\bf x}) +\lambda^{\dag}\sum_{\bf n}\langle -\widehat\beta ({\bf x}),
\phi_{\bf n} \rangle \phi_{\bf n} /
(\lambda_{\bf n} - \lambda^{\dag})\big)=
 \eqno{(24)}
$$
$$
=\lambda^{\dag} \sum_{\bf n}\frac{-\widehat\beta ({\bf x}),\phi_{\bf n} \rangle 
\phi_{\bf n}}{\lambda_{\bf n}}+
(\lambda^{\dag})^2 \sum_{\bf n}  
\frac{\langle -\widehat\beta ({\bf x}),\phi_{\bf n} \rangle 
\phi_{\bf n}}{\lambda_{\bf n}(\lambda_{\bf n} - \lambda^{\dag})}=
$$
$$
\lambda^{\dag}\sum_{\bf n}
\frac{\langle -\widehat\beta ({\bf x}),\phi_{\bf n} \rangle 
\phi_{\bf n}}{\lambda_{\bf n}-\lambda^{\dag}}=\phi+\widehat\beta ({\bf x}).
$$
We represent $\phi({\bf x})$ (solution of the inhomogeneous equation) using (Eq. 24) as:
$$
\phi ({\bf x}) = -\widehat\beta ({\bf x}) +\lambda^{\dag}\sum_{\bf n} 
\langle -\widehat\beta ({\bf x}),\phi_{\bf n} \rangle \phi_{\bf n}\lambda_{\bf n}^{-1} + 
(\lambda^{\dag})^2
$$
$$
\sum_{\bf n}
\langle -\widehat\beta ({\bf x}),\phi_{\bf n} \rangle 
\phi_{\bf n}\lambda_{\bf n}^{-1}(\lambda_{\bf n}-\lambda^{\dag})^{-1} =
$$
$$
=  -\widehat\beta ({\bf x}) +
$$
$$
\lambda^{\dag} \int_\Omega
\underbrace{
\bigg( {\mathcal K}({\bf x},{\bf x}') + 
\lambda^{\dag} 
\sum_{\bf n} \phi_{\bf n}({\bf x}) \phi_{\bf n}({\bf x}')
\lambda_{\bf n}^{-1}(\lambda_{\bf n}-\lambda^{\dag})^{-1}
\bigg)}_{\Gamma ({\bf x},{\bf x}',\lambda^{\dag})}
$$
$$
(-\widehat\beta ({\bf x}))d{\bf x}'.
\eqno{(25)}
$$

Thus, the solution of the inhomogeneous Fredholm equation for $\lambda^{\dag}\neq \lambda_{\bf n}$
expressed using the resolvent $\Gamma ({\bf x},{\bf x}',\lambda^{\dag})$  for the integral kernel $-|{\bf x}-{\bf x}'|^{-1}$ (this representation of the solution, generally speaking, is valid also in the case of deviation from the spherical symmetry of the problem).
For the case $\lambda^{\dag}= \lambda_{\bf n}$, there is no solution to the inhomogeneous equation, 
since
$\langle \widehat\beta ({\bf x}),\sin({\bf n}{{\bf x}}) \rangle \neq 0$ 
(according to Fredholm's $3$rd theorem).

For an integral kernel that differs from the classical Newtonian potential,
a rep\-re\-sen\-ta\-ti\-on of the solution of the Fredholm equation in terms of a resolvent with sinusoidal 
eigenfunctions
is impossible. We go on to demonstrate the technique for constructing the resolving resolvent
for semi-symmetric integral Schmidt--type kernels.

We  now consider the connection between the criterion for the 
implementation of periodic solutions
and the previously introduced Helmholtz equation. 
The equation can be represented as the following expansion
$$
\int {\mathcal K}( |{\bf x}-{\bf x}'|)\phi ({\bf x}')d{\bf x}'=\sum_{j=0}\xi_{2j}
d^{2 j}\phi({\bf x}) /d {\bf x}^{2j},
\eqno{(26)}
$$ 
\noindent
where 
$$
\xi_{2j}=\int{\mathcal K}(|{\bf x}-{\bf x}'|)
|{ {\bf x}}-{{\bf x}}'|^{2j}d{\bf x}',
\eqno{(27)}
$$ 
\noindent
represents the even moments of interaction energy.
Then the previously presented homogeneous Fredholm equation of the second kind
can be reduced (in the case ${j=0, 1}$)
to the Helmholtz equation: 
$$
\Delta \phi + \lambda_H \phi =0,~~~
\lambda_{H}=(1+\lambda_{II}\xi_0)/(\lambda_{II} \xi_{2}).
\eqno{(28)}
$$
The solutions of this equation are spherical waves decreasing in amplitude, as follows:
$$
\sim r^{- 1}\sum_{j=1}^2{\mathfrak a}_j
\exp (i\sqrt {\lambda_{H }}r ),
\eqno{(29)}
$$
\noindent
with a spatial period of: 
$$
T =2\pi {\lambda_H }^{-1/2}=
2\pi \sqrt {\lambda_{II}\xi_{2}/(1+\lambda_{II}\xi_0 )}.
\eqno{(30)}
$$
Moreover, the admissible size of the region with the Dirichlet boundary conditions 
can be unlimited;
in this case, assuming the Sommerfeld conditions, we can obtain
the only solution to the problem for a gravitational source (including the right-hand-side of 
the Helmholtz equation) with the conditions
at infinity as follows:
$$
\phi(r)\sim\int c^2\Lambda/(r\theta)\exp(i\sqrt{\lambda_{H}}r)d{r}.
\eqno{(31)}
$$
However, if we accept that it is possible to set
spatial periodic conditions corresponding to the existence of 
massive multi-particle systems,  then the solution of such a problem
should be considered in the form of
superpositions of divergent and convergent spherical waves from adjacent sources.
At the same time, an interesting problem of interpretation in the physical
space of the results of interference of these waves, including quasi-two-dimensional formations
with increased density, is arising due to the implementation of conditions for
occurrence of wave beats.

A similar situation arises in the plane case ($d=2$) when the solutions of the Helmholtz equation
are representable in terms of   zeroth--order Hankel functions  \citep{Svesh}:
$$
\phi_2 =\sum_{j= 1}^2
{\mathfrak b}_j H_0^{(j)},
\eqno{(32)}
$$\noindent
with a change of semi-periodicity vs monotonic decrease.

A nonlinear Hammerstein equation (Eq. 13) with symmetric kernel and quadratic
nonhomogeneous term satisfies the existence conditions for the
 $(\lambda_0, U_0({\bf x} )) $ solutions \citep{Reg}. We can then consider the possibility of constructing a locally unique
solutions of (Eq. 13)
with a nearly
non-trivial solution of the inhomogeneous linear equation obtained above (on a ``plateau''
corresponding to the
cosmological self-consistent gravitational potential.
To do this, we apply the   Bratu perturbation method, 
introducing new variables $\delta\lambda, \delta U$
($\lambda=\lambda_0+\delta\lambda$, $U({\bf x})=U_0({\bf x})+\delta U({\bf x})$
and we rewrite the equation (Eq. 13) accordingly:
$$
\delta U({\bf x})=\delta\lambda\int_{\Omega'} {\mathcal K}(|{\bf x}-{\bf x}'|)
\sum_{j\ge 0}\frac{(\delta U)^j}{j!}(\exp(-U))^{(j)}\big|_{U=U_0}d{\bf x}'+
\eqno{(33)}
$$
$$+
\lambda_0\int_{\Omega'} {\mathcal K}(|{\bf x}-{\bf x}'|)
\sum_{\ell \ge 1}\frac{(\delta U)^\ell}{\ell!}(\exp(-U))^{(\ell)}\big|_{U=U_0}d{\bf x}',
$$
$$
U_0=\lambda_0\int_{\Omega'} {\mathcal K}(|{\bf x}-{\bf x}'|)\exp(-U_0)+\alpha{\bf x}^2.
$$
Representing the solution as a series in powers of a small parameter
$\delta U( {\bf x})=\sum_{j\ge 1} h_j (\delta \lambda )^j$ and substituting into the equation,
we obtain a (infinite) system of equations for determining the coefficients:
$$
h_1({\bf x})=
\lambda_0\int{\mathcal K}(|{\bf x}-{\bf x}'|)
(-\exp(U_0({\bf x}')))h_1({\bf x}')d{\bf x}'+
$$
$$
\int{\mathcal K}(|{\bf x}-{\bf x}'|)
\exp(U_0({\bf x}'))d{\bf x}',...,
\eqno{(34)}
$$
$$
h_j({\bf x})=
\lambda_0\int{\mathcal K}(|{\bf x}-{\bf x}'|)
(-\exp(U_0({\bf x}')))h_j({\bf x}')d{\bf x}'+\mu_j({\bf x}),~~~
\eqno{(35)}
$$
$$
|\mu_j({\bf x})|<P_1\cdot
 {\rm{max}}_{\mu_j}\big(|\lambda_0|(\mu_{j,2}/2!+
$$
$$\mu_{j,3}/3!)+...+\mu_{j,j}/j!)+
(\mu_{j-1,1} +\mu_{j-1,2}/2!+...+ \mu_{j-1,j-1}/(j-1)!)  \big),
\eqno{(36)}
$$
$$
|h_j({\bf x})|<|\mu_j({\bf x})|(1+|\lambda_0|P_2)<P_3,
$$
$$
\mu_{j,s}=\sum\mu_{i_1}...\mu_{i_s},~i_1+...+i_s=j,~~i_{1,2,...}\in\{1,2,...,j\}.
\eqno{(37)}
$$
\noindent

The above   representation of $\delta U( {\bf x})$ as a regular series
in the neighborhood of $U_0({\bf x })$ due to the presence of majorants of the coefficients
only. However, if the value of the Fredholm determinant 
for the linearized equation
 $D(\lambda )\to 0$ (at the point $(\lambda_0,U_0$)), then 
the second branch of the Hammerstein equation 
solution becomes relevant.

Thus, we get an analytical representation of the solution
the Hammerstein equation for the potential, which is unique
by construction and
whose existence can be guaranteed in some neighborhood of
equilibrium solution, $U_0$. 
This means the resistance of the system to external influences (the presence of external gravitational fields,
varying within a certain spatial region) and final changes in the most self-consistent
system field. The solutions of the linearized and nonlinear equations for the potential,
define of a quasi-periodic nature of the structures, when the difference in the amplitudes of the local 
maxima of the potential
depends on the relation $\alpha /\lambda_{II}$. If we accept that extrema
density distribution of matter correspond to nodes and anti-nodes on
solutions of equation (Eq. 13), it becomes obvious that one-dimensional (1D) and two-dimensional (2D) structures,
observable in cosmological scales can be explained as completely deterministic objects,
without the involvement of stochastic dynamics.

\section{Linearized equations of general form and construction of a new branch of a nonlinear equation}

Above, we consider the case of Fredholm equations linearized near the point of constant potential. 
In doing so, we assumed that the kinetic temperature parameter
has a fixed value. Clearly, there is a question of clarification
of the equilibrium function and of the subsequent behavior of the system of gravitating particles.

We assume that the Hammerstein equation (Eq. 13) has a local equilibrium solution of
  $U=U_0({\bf x})$ and we consider a small deviation from it, $\varphi ({\bf x})=U-U_0$.
  For this, the difference of two equations of the form (13) is usually considered as:
  $$
  (U_0+\varphi)-U_0=\lambda \big(\widehat{\mathfrak G}({U_0+\varphi})- \widehat{\mathfrak G}({U_0})\big),
\eqno{(38)}
  $$
\noindent
where
we obtain the (homogeneous) Fredholm equation of the second kind, which takes the following form:
  $$
  \varphi({\bf x})+
  {\lambda}\int_\Omega
  \underbrace{{\mathcal K}(|{\bf x}-{\bf x}'|)
  \exp\big(-U_0 ({\bf x}') \big)}_{{\mathcal L}({\bf x},{\bf x}')}
  \varphi({\bf x}')d{\bf x}'=0.
  \eqno{(39)}
  $$
We consider the last equation, including (as before)
  the inhomogeneous term $\alpha{\bf x}^2$. This corresponds to the replacement of the nonlinear integral
  operator with
its Frechet derivative (near the solution $U_0$). Generally speaking,
to interpret the statement of the problem of inclusion in the consideration in a linear approximation, we can
either assume: 1)   the equilibrium state $U_0({\bf x})$ corresponds to
homogeneous nonlinear equation and the presence of an inhomogeneous right-hand-side
a priori is due to solutions-deviations from $U_0$; 2) or it is possible in a nonlinear
Hammerstein equation to change the variable $U^\dag=U-\alpha {\bf x}^2$
and at the same time to modify the integral kernel appropriately
$$
{\mathcal K}(|{\bf x}-{\bf x}'|)\to
{\mathcal K}^\dag(|{\bf x}-{\bf x}'|)\equiv {\mathcal K}(|{\bf x}-{\bf x}'|)
\exp\big( -\alpha {\bf x}^2 \big).
  \eqno{(40)}
$$
Then Eq. 33 can be tranformed to a linear equation for $U^\dag ({\bf x})$,
whose kernel will contain information about the inhomogeneity of the original Hammerstein equation.

The  discussion that follows mainly refers to the first variant of accounting for inhomogeneity,
but it can be extended in a completely trivial way to the second option (if
we are further taking $U_0\to U_0-\alpha {\bf x}^2$).

The kernel of the integral ${\mathcal L}({\bf x},{\bf x}')$ belongs to the class of
Schmidt kernels and can be written in the following form:
$$
{\mathcal L}({\bf x},{\bf x}') =
\Upsilon ({\bf x}',{\bf x})
{\mathcal K}(|{\bf x}-{\bf x}'|)
$$
$$
\sqrt{\exp\big(-U_0 ({\bf x})\big)
\cdot \exp\big(-U_0 ({\bf x}') \big)}=
  \eqno{(41)}
$$
$$
=
   \Upsilon ({\bf x}',{\bf x})
  {\mathcal K}^\dag({\bf x},{\bf x}'), ~~~
  \Upsilon ({\bf x}',{\bf x})\equiv
  \sqrt{\frac{\exp\big(-U_0 ({\bf x}') \big)}{\exp\big(-U_0 ({\bf x}) \big)}}.
$$
We use $\Gamma ({\bf x},{\bf x}', {\lambda})$ to denote the resolving kernel
for the (weakly polar) integral kernel  
${\mathcal K}({\bf x},{\bf x}')$, whose expression for it was obtained in Section 2. Thus
the resolvent satisfies the functional equation:
$$
\Gamma ({\bf x},{\bf x}', {\lambda}) = {\mathcal K}({\bf x},{\bf x}') +\lambda\int_\Omega
({\bf x},{\bf x}'') \Gamma ({\bf x}'',{\bf x}', {\lambda})d{\bf x}''.
  \eqno{(42)}
$$
If we set:
$$
\Gamma^\dag ({\bf x},{\bf x}', {\lambda}) =
(\Upsilon ({\bf x}',{\bf x}))^{-1}
\Gamma ({\bf x},{\bf x}', {\lambda}),
  \eqno{(43)}
$$
\noindent
then we have:
$$
\Gamma^\dag ({\bf x},{\bf x}', {\lambda}) = {\mathcal L}({\bf x},{\bf x}') +\lambda\int_\Omega
{\mathcal L}({\bf x},{\bf x}'') \Gamma^\dag ({\bf x}'',{\bf x}', {\lambda})d{\bf x}'',~~
  \eqno{(44)}
$$
$$
{\mathcal K}^\dag({\bf x},{\bf x}'')\Gamma^\dag ({\bf x}'',{\bf x}', {\lambda})=
\Upsilon ({\bf x}',{\bf x}){\mathcal K}({\bf x},{\bf x}'')\Gamma ({\bf x}'',{\bf x}', {\lambda}).
  \eqno{(45)}
$$
The resolvent kernel corresponding to the integral kernel
${\mathcal K}({\bf x},{\bf x}')\exp\big(-U_0({\bf x}')\big)$
will be the product $\Upsilon ({\bf x}',{\bf x}) \Gamma^\dag ({\bf x},{\bf x}', {\lambda})$,
where $\Gamma^\dag ({\bf x},{\bf x}', {\lambda})$ is in 
resolution for the integral kernel ${\mathcal K}^\dag({\bf x},{\bf x}')$.

By construction, for the ${\mathcal L}$ kernel, we use
sinusoidal oscillations with space-dependent
variable coefficients
as eigenfunctions, which means that when constructing a solution to 
a compositional solution, we can observe a superposition of 
a function close to a two-particle potential with repulsion of Eq.(1) 
and oscillations with a weakly varying amplitude.

It should be noted that based on the structure of solution (14), as well as the corresponding solution
of the linearized equation
for ${\mathcal K}\to {\mathcal L}$, we can see that the presence of distinguished
directions in expansions of the anisotropy of the vectors,
$$
(0,0,j)_{j=1,...,n},~~~
(0,j_1,j_2)_{j_1=1,...,J_1;\,j_2=1,...,J_2}),
\eqno{(46)}
$$
in the background of a homogeneous cosmological expansion and leads to the formation
1D and 2D large structures. We note that this mechanism drastically
differs from those of Zeldovich pancakes, since it is completely deterministic
and has nothing to do with stochastic perturbations.

The form of solving the linearized equation obtained above can be used
for
construction of analytic branches in the neighborhood of the above solution, and of a possible
new branch of the Hammerstein equation ($D(\lambda)\to 0$).
If $\lambda_0$ is not a kernel eigenvalue
${\mathcal L}({\bf x},{\bf x}')$, then in the neighborhood of $\lambda_0$
of the original Hammerstein equation has a unique
  holomorphic (with respect to $\lambda-\lambda_0$) solution $U_0({\bf x},\lambda)$,
  tending toward $U_0(\bf x)$ as $\lambda \to \lambda_0$.
  
        The original equation can be written as:
  $$
  U_0 + \phi({\bf x})=(\lambda_0+\delta\lambda)\int {\mathcal L}_+({\bf x},{\bf y})
  \exp(U_0)\big(1-\phi({\bf x}')+
        $$
        $$
        (1/2!)\phi({\bf x}')-...
  \big)d{\bf x}',~~{\mathcal L}_+={\mathcal L}\exp(-\alpha {\bf x}^2).
  \eqno{(47)}
  $$
We represent the function $\phi ({\bf x})$ as a series:
$$
\phi ({\bf x}) =(\delta\lambda) \phi_1({\bf x})+ (\delta\lambda)^2 \phi_2({\bf x})+
...,
\eqno{(48)}
$$
\noindent
where
$$
\phi_m ({\bf x})= \psi_m(\phi_1,...,\phi_{m-1};\,{\bf x}) +\lambda_0
\int\Gamma({\bf x},{\bf x}',
$$
$$
\lambda_0)
\psi_m(\phi_1,...,\phi_{m-1};\,{\bf x}')d{\bf x}'.
\eqno{(49)}
$$
The branching condition for the equilibrium solution $U_0({\bf x})$ 
is the presence of $\lambda_0$ an eigenvalue of the kernel ${\mathcal L}_{+}$,
as well as the following condition:
$$
\int_\omega \exp(U_0({\bf x}'))\phi_1({\bf x}')d{\bf x}'\neq 0.
\eqno{(50)}
$$
In this case, we need to consider the Puiseux series in powers of
$(\delta\lambda)^{1/2}$
$$
\phi ({\bf x})=(\delta\lambda)^{1/2}\phi_1({\bf x})+
(\delta\lambda)^{2/2}\phi_2({\bf x})+...
\eqno{(51)}
$$
Then, the considered Hammerstein equation has: 1) two different
solutions corresponding to one $\lambda\ge \lambda_0$ and none
solution corresponding to $\lambda<\lambda_0$,
provided that $\langle \exp(-U_0),\phi_1 \rangle$
and $\langle \lambda_0 \exp(-U_0), \phi_1^3 \rangle$
have different signs; 2) two solutions for $\lambda<\lambda_0$
and none for $\lambda >\lambda_0$,
if $\langle \exp(-U_0),\phi_1 \rangle$
and $\langle \lambda_0 \exp(-U_0), \phi_1^3 \rangle$
have the same signs (in the system we are considering, conditions must arise for the implementation of the first option).

Thus, we carried on with the following steps. Although the linear equation (Eq. 14) does not contain an inhomogeneity, it carries information about the Dirichlet conditions and contains Green’s function \citep{Now}
associated with the original Poisson equation. So,
we considered the inhomogeneous linear equation taking into
account quadratic repulsion $\alpha (\theta, \Lambda){\bf x}^2$.
The use of an integral equation of the second kind for solving the internal Dirichlet problem is due to
the presence of inhomogeneous boundary conditions associated with this additional term growing with increasing distance, as follows from the theorem \citep{G1} on the general form of the gravitational potential of a sphere and a point.
Therefore, we use the representation of the solution of the internal problem
in the form of a double layer potential \citep{Now}, as considered in detail in \citep{GFC2}, as per Eqs. 26-29.
The integral representation of the solution of the Dirichlet problem leads us to an equation Eq.(11), then via Hammerstein equation, to its linearized form of the Fredholm equation.

For the homogeneous Fredholm equation of the form given in Eq. 14 
there is a countable system of fundamental functions in the general form of spherical harmonics \citep{Atk}.
These functions correspond to the fundamental values 
$\lambda (=\lambda_{\bf n})$, which are (specifically, degenerate, generally speaking,   when the non-sphericity of the problem is taken into account)
roots of the algebraic equation $D(\lambda)=0$, where $D(\lambda)=1+\sum_{j=1}^{\infty} (-1)^j A_j/j!$  is the Fredholm determinant
of the kernel ${\mathcal K}$  \citep{Zem}; coefficients $A_j$ are expressed in terms of the
integral of the determinants of a symmetric matrix with elements
${\mathcal K}({\bf x}_k,{\bf x}_\ell)\big|_{k,\ell=1,...,j}$.
In accordance with the first Fredholm theorem \citep{Waz}, the solution of an inhomogeneous second kind equation   is unique (of class $C \cup L_2$) for values of the 
parameter $\lambda$ that do not coincide with the set of roots
$D(\lambda)$; moreover (for $D(\lambda)\neq 0$), in accordance with the Fredholm alternative,
the corresponding homogeneous equation has only a trivial solution.
Then, the construction of a solution to the inhomogeneous Fredholm equation is obtained
using a (resolvent) meromorphic function $D({\bf x},{\bf x}';\lambda)/D(\lambda)$, where $D({\bf x},{\bf x}' ;\lambda)$ is the first minor of the Fredholm determinant.

 \section{Conclusions}

We consider the emergence of matter structures in the Local Universe based on the Vlasov kinetic technique.
We show a deterministic mechanism of structure formation that is due to the self-consistent gravitational field of particles on local scale,
as distinct to the stochastic mechanism due to the evolution of the primordial density perturbations within pancake theory on global scale.
Hence, within this approach the local (late) and global (early) Universe reveal different mechanisms of structure formation. The crucial aspect of our analysis is the consideration of the modified gravitational interaction given by Eq. 1 with a repulsive cosmological term, which is based on the theorem \citep{G1} of the identity of the sphere and point mass gravitational fields (however, with the non-force-free field within a shell). 

Our analysis includes the transformation of the Poisson equation to Liouville-Gelfand equation, then a hierarchy of Dirichlet problems
for differential equations of Helmholtz type. 
The transition from differential equations for gravitation to integral equations of the Hammerstein type is considered
through bulk density.  The resulting inhomogeneous Fredholm integral equation
of the second kind was analysed using the eigenfunctions of the weakly polar kernel of the homogeneous equation. The solutions
were found through the resolvent predicting the formation of 1D and 2D filamentary semi-periodic structures as a result of self-consistent gravitational interaction involving the cosmological constant.
Methodically, the novelty of our analysis is in the use of inhomogeneous Fredholm integral equation of the second kind for solving the internal Dirichlet problem, with inhomogeneous boundary conditions and with an additional term growing with distance following from sphere-point identity theorem. 

Observationally, the characteristic scale of the considered structure formation mechanism is given by the critical radius $ r_{crit}= \big(3G M/\Lambda c^2\big)^{1/3}$, defining the mutual role of the attracting and repulsive terms in Eq. 1. This implies that the sizes of the voids  can vary depending on the local region (e.g., \citep{spot,S1,S2}) and local galactic flows can occur. To illustrate, let us consider the case of the Virgo cluster: the mass is $1.2 \times 10^{15} M_{\odot}$ within the radius $2.2$ $Mpc$, the center of the cluster is located in $16.5 \pm 0.1$ $Mpc$ from us \citep{VC,Mei,Ka}, then $r_{crit}= 11.80$ $Mpc$. This implies, that since the $r_{crit}$ exceeds the distance of the Local Group from the center of the Virgo cluster the $\Lambda$-term is able to describe their observed mutual repel, as a local H-flow, in entire accordance with the reported value of the Hubble constant.
Thus, the cosmological constant in the potential determines not only the appearance of the structures but also defines their scale, for instance, the sizes of the voids depending on the mean density of the local region and hence differing for various regions.

The principal conclusion of this work is that the stochastic mechanism of structure formation can be aptly replaced by deterministic one in local regions of the Universe. Along with the Hubble tension, this can be another indication for intrinsic differences in the early and late Universe.

\section{Acknowledgments}

We are thankful to the referee for valuable comments. VMC is acknowledging the Russian Science Foundation grant 20-11-20165.

\end{document}